\documentclass[12pt]{article}
\usepackage{amsfonts}

\makeatletter\@addtoreset{equation}{section}\makeatother

\def\bR {\mathbb{R}}

\input epsf

\newcommand{\figin}[2]{
\begin{figure}[t]
\centerline{\hbox{\epsffile{#1.eps}}}
\centerline{\parbox{12cm}{\caption{#2\label{#1}}}}
\end{figure}}

\newcommand{\vev}[1]{{\left< {#1} \right>}}

\newcommand{\Tr}{{\rm Tr\,}}

\newcommand{\cF}{{\cal F}}

\newcommand{\cN}{{\cal N}}

\newcommand{\preprint}[1]{\begin{table}[t]  %%
             \begin{flushright}               %%
             {#1}                             %%
             \end{flushright}                 %%
             \end{table}}                     %%
\renewcommand{\title}[1]{\vbox{\center\LARGE{#1}}\vspace{5mm}}
\renewcommand{\author}[1]{\vbox{\center#1}\vspace{5mm}}
\newcommand{\address}[1]{\vbox{\center\em#1}}
\newcommand{\email}[1]{\vbox{\center\tt#1}\vspace{5mm}}

\begin{document}

\begin{titlepage}
\preprint{hep-th/0501109 \\ 
ITFA-2005-01
}

\title{All-genus calculation of Wilson loops using D-branes}

\author{Nadav Drukker$^1$ and Bartomeu Fiol$^2$}

\address{$^1$The Niels Bohr Institute, Copenhagen University\\
Blegdamsvej 17, DK-2100 Copenhagen, Denmark\\
$^2${Institute for Theoretical Physics, University of Amsterdam\\
1018 XE Amsterdam, The Netherlands}}

\email{drukker@nbi.dk, bfiol@science.uva.nl}

\abstract{
The standard prescription for calculating a Wilson loop in the 
$AdS$/CFT correspondence is by a string world-sheet
ending along the loop at the boundary of $AdS$. For a multiply 
wrapped Wilson loop this leads 
to many coincident strings, which may interact among themselves. 
In such cases a better description of the system is in terms of a D3-brane 
carrying electric flux. We find such solutions for the single straight 
line and the circular loop. The action agrees with the string 
calculation at small coupling and in addition captures all the 
higher genus corrections at leading order in $\alpha'$. The resulting 
expression is in remarkable agreement with that found from a zero 
dimensional Gaussian matrix model.
}

\end{titlepage}

\section{Introduction}

Some of the most interesting observables in gauge theories are Wilson loop 
operators, the holonomy of the gauge field around a contour. The 
expectation value of these operators gives the effective action for 
a charged particle following that path. A hallmark for confinement 
is the area-law of the Wilson loop---when its expectation value is 
proportional to the area enclosed.

In this paper we do not deal with a confining theory, rather 
with a conformal gauge theory, maximally supersymmetric 
($\cN=4$) Yang-Mills, which has a dual description 
at strong coupling via string theory on $AdS_5\times S^5$ 
\cite{ads/cft}. Soon after the proposal of this duality the 
prescription for calculating Wilson loops in string theory was put forth 
by Rey and Yee and by Maldacena \cite{rey-wl,maldacena-wl}. 
The prescription is to consider a fundamental 
string ending on the boundary of $AdS$ along the path specified by 
the Wilson loop operator.

This suggestion is very intuitive, after all the hope for a string 
description of confining theories relies on the area of the string bounded 
by the Wilson loop providing the area law. In the case of the conformal 
theory the peculiarities of the $AdS$ geometry give an action that is 
not proportional to the area, but rather is scale independent.

Since its inception, this has been the standard method of evaluating 
Wilson loop operators in all the generalizations of the $AdS$/CFT 
correspondence, including to theories in other dimensions and to 
confining theories. But here we would like to propose an alternative 
way of evaluating the Wilson loop operator with D3-branes rather 
than fundamental strings. One inspiration for this 
is the concept of ``giant gravitons,'' where fundamental string excitations 
are replaced by spherical D3-branes wrapping part of the $S^5$ 
\cite{McGreevy:2000cw} or part of $AdS_5$ 
\cite{Grisaru:2000zn,Hashimoto:2000zp}. Is there an analogous 
effect for the big strings that describe Wilson loops?

The idea of describing a fundamental string in terms of a D3-brane is 
actually not new, and to find our solution we follow closely the 
construction of Callan and Maldacena \cite{Callan} 
(see also \cite{Gibbons}). There 
they find a solution to the equations of a D3-brane in flat space in which 
the brane has a localized spike. This spike is analogous to a string 
ending on the D3-brane. Here we adapt their calculation to the $AdS$ 
background and replace the fundamental string describing the Wilson 
loop with a D3-brane.

The description of the Wilson loop in terms of a fundamental string is 
a well established part of the $AdS$/CFT dictionary. So what is the 
role of D3-brane solutions we find below? The geometry of the branes 
will be such that they pinch off at the boundary of $AdS$, ending 
along the curve defined by the Wilson loop observable. The branes will 
carry electric 
flux, which is the same as fundamental string charge, so they can play 
the same purpose as the fundamental string themselves. In fact, 
already in one of the original papers on Wilson loops in the $AdS$/CFT 
correspondence \cite{rey-wl} it was shown that D3 branes can also be 
used for that purpose.

We will be particularly interested in the case when the Wilson loop 
is described by a large number of fundamental strings. This happens 
when the operator involves many coincident Wilson loops, a multiply wound 
Wilson loop, or a Wilson loop in a high-dimensional representation. 
Those cases differ by the trace structure in the gauge theory, or the 
connectivity of the string surfaces \cite{Gross:1998gk}. In 
all three cases the leading planar behavior should be the same and scale 
like the multiplicity of the loop $k$. The subleading behavior is an 
interesting question, that resembles a similar issue in confining theories,
as QCD. For confining theories, the flux tube connecting $k$ quarks and $k$ 
anti-quarks is called a $k$-string, and its tension $\sigma_k$ is not just
$k\sigma_1$. {\it E.g.}, for softly broken ${\cal N}=2$ SYM, Douglas and
Shenker \cite{douglas} found
\begin{equation}
\sigma_k=N\Lambda^2\sin \frac{\pi k}{N}=\Lambda ^2 \pi k-
\frac{\Lambda ^2\pi^3}{3!}\frac{k^3}{N^2}+\dots
\end{equation}
and subsequently this formula has also appeared in MQCD \cite{hanany} and in 
supergravity duals of ${\cal N}=1$ SYM \cite{herzog}. We will compare the 
scaling of both subleading terms in the conclusions. 

This subleading behavior may be accounted for by the interaction among 
the $k$ string surfaces, so it requires to study this system beyond the 
planar approximation. This is a complicated question and no exact results 
have ever been calculated. The D3-brane solutions we construct will 
provide a shortcut to finding all those non-planar contributions!

The reason for this already appeared in \cite{Callan}. It 
was argued that the system of many coincident strings is better 
described in terms of the dynamics of the D3-brane they end on. 
Applying that logic to the Wilson loop calculation we see that the 
non-planar contributions, which become important for the multiply 
wrapped loops should be captured by the D3-brane dynamics.

We will study the two simplest Wilson loop observables, the infinite 
straight line and the circle. 
Let us first review the standard calculation of those two Wilson loops. 
The supersymmetric Wilson loop (in a space of Euclidean signature) is
\begin{equation}
W=\frac{1}{N} \Tr {\cal P} \exp i\oint (A_\alpha \dot x^\alpha
+i\Phi _i |\dot x| \theta ^i)dt\,,
\end{equation}
where $A_\alpha$ is the gauge field, $\Phi_i$ the six scalars, 
$x^\alpha(t)$ parameterizes a path in space and we take $\theta^i$ to be 
a constant unit vector in $\bR^6$.

The simplest Wilson loop imaginable is a single infinite line. Its 
expectation value corresponds to the exponential of the renormalized 
mass of the probe particle times the length of the line. In the case of 
$\cN=4$ there should be no renormalization of the mass, and the 
expected result is simply $\vev{W_{\rm line}}=1$. This can indeed be 
shown both in the gauge theory and string theory, and is a consequence 
of this operator preserving half of the supercharges.

To be specific, consider a Wilson loop extended in the $x^1$ direction and 
localized in the transverse directions $x^2=x^3=x^4=0$. We use 
the coordinate system for $AdS_5$
\begin{equation}
ds^2=\frac{L^2}{y^2}
\left(dy^2+(dx^1)^2+(dx^2)^2+(dx^3)^2+(dx^4)^2\right)\,.
\end{equation}
Here $L$ is the curvature radius of the space, related to the 
string coupling $g_s$, string length $l_s$ and the 't Hooft coupling 
$\lambda=g_{YM}^2N$ by 
$L=(4\pi g_s N)^{1/4}l_s=\lambda^{1/4}l_s$.

The standard description of the Wilson loop ending along the line is 
the surface spanned by the coordinates $y$ and $x^1$ at 
$x^2=x^3=x^4=0$. The 
area of this surface (calculated with the induced metric) will have a 
divergence from small $y$, and will be proportional to the length 
of the $x^1$ direction, $X^1$. With a cutoff $y_0$ the regularized 
area is
\begin{equation}
A=\sqrt\lambda\frac{X^1}{2\pi y_0}\,.
\end{equation}
As explained in \cite{hirosi} (and reviewed below) one should 
add a boundary term that cancels this divergence. Then the full action 
is $S_{\rm string}=0$, and indeed the expectation value of the 
Wilson loop is simply unity.

Next we consider the circular Wilson loop 
\cite{Berenstein:1999ij,hirosi}, where the path follows a circle 
of radius $R$ in the $x^1,x^2$ plane. 
The string solution describing this loop 
(using polar coordinates) is given by 
$r^2+y^2=R^2$ and the bulk action is
\begin{equation}
S_{\rm bulk}
=\sqrt\lambda\int_{y_0}^R dy\,\frac{r}{y^2}\sqrt{1+r'^2}
=\sqrt\lambda\int_{y_0}^R dy\,\frac{R}{y^2}
=\sqrt\lambda\left(\frac{R}{y_0}-1\right)\,,
\label{circle-string}
\end{equation}
The divergence is removed by the boundary term and we are left with 
the answer
\begin{equation}
\vev{W_{\rm circle, string}}=\exp\sqrt\lambda\,.
\end{equation}

The final answer is independent of $R$, due to conformal invariance.
The circular Wilson loop is related to the straight line by a conformal 
transformation, so it also preserves half the supersymmetry
\cite{Bianchi:2002gz,Mikhailov:2002ya}, yet its expectation value is 
not unity. For the straight line the combined gluon and scalar propagator 
vanishes, while for the circle it's a finite constant. This allowed Erickson, 
Semenoff and Zarembo \cite{erickson} to sum all rainbow and 
ladder diagrams and their calculation reproduced the above 
$\exp\sqrt\lambda$ result.

The reason for the difference between the line and the circle is rather 
subtle. In applying the conformal transformation 
mapping the line to the circle one needs to add the point at infinity to 
the line. Consequently there is a slight difference that in perturbation 
theory manifests itself as an addition of a total derivative to the propagator 
\cite{Drukker:2000rr}. Under 
the assumption that this modification of the propagator is the only change, it 
was shown there how to write the expectation value of the 
Wilson loop in terms of zero dimensional Gaussian matrix model.

This matrix model was solved exactly and written as an asymptotic 
expansion to all orders in $1/N$ and $1/\sqrt\lambda$. As was 
mentioned, the leading planar result of the matrix model, 
$\exp\sqrt\lambda$ agrees with the string calculation in $AdS$. 
In our case the D3-brane calculation will reproduce this term 
correctly, as well as an infinite series of corrections of the form 
$\lambda^{k+1/2}/N^{2k}$. All those terms will be in precise 
agreement with the perturbative non-planar calculation as given 
by the matrix model!

The paper is organized as follows. We start with the simple example of the 
infinite straight line in the next section, calculating the action, explaining 
the necessary boundary terms and proving supersymmetry. In Section 3 we 
turn to the richer case of the circular loop and compare the results to 
the matrix model. Next we describe how to use D3-branes to calculate 
't Hooft loops and comment also on the perturbative calculation of those 
observables. We end with some discussion.

\section{Infinite straight line}

As a warm up to explore our idea, we look first at the infinite straight 
single Wilson line in $\bR^4$. It is extended in the $x^1$ direction and 
localized in the transverse directions $x^2=x^3=x^4=0$. 
This operator exists in both Euclidean and Lorentzian signature, 
and our construction will work perfectly well in both. We use Euclidean 
conventions to be consistent with the other example which we study, 
the circle. We will switch briefly to Lorentzian signature to prove 
that the solution is supersymmetric. The description of the infinite 
straight Wilson loop in terms of a D3-brane was originally done by 
Rey and Yee \cite{rey-wl}.

\subsection{D3-brane solution}

To study this Wilson loop it is helpful to use spherical coordinates in 
the directions transverse to the line, which we choose lo lie along 
$x_1$, so for $AdS_5$ we use the coordinate system
\begin{equation}
ds^2=\frac{L^2}{y^2}
\left(dy^2+(dx^1)^2+dr^2
+r^2d\Omega_2^2\right)\,.
\end{equation}

Here we want to reproduce $\vev{W}=1$, where instead of a fundamental 
string we use a D3-brane carrying electric flux. This D3-brane will be 
a hypersurface in $AdS_5$ given by a single equation, which by the 
symmetries of the problem is clearly $y=y(r)$. So we use $x^1$, $r$, 
$\theta$ and $\phi$ as world-volume coordinates and turn on 
the scalar field $y(r)$ as well as the electric field $F_{1r}(r)$.

The action includes the Dirac-Born-Infeld (DBI) part and the 
Wess-Zumino (WZ) 
term, which captures the coupling to the background Ramond-Ramond 
fields. In $AdS_5$ 
there are $N$ units of flux of the Ramond-Ramond five-form, whose 
potential can be taken to be
\begin{equation}
C_4=\frac{L^4r^2\sin \theta}{y^4} dx^1\wedge dr
\wedge d\theta \wedge d\phi\,.
\label{C4}
\end{equation}

We thus find the action
\begin{eqnarray}
S&=&T_{\rm D3}\int e^{-\Phi}\sqrt{\det(g+2\pi\alpha' F)}
-T_{\rm D3}\int P[C_4]
\nonumber\\
&=&\frac{2N}{\pi}\int dx^1 dr\, 
\frac{r^2}{y^4}\left(
\sqrt{1+y'^2+(2\pi \alpha' F_{1r})^2 \frac{y^4}{L^4}}-1 \right ).
\label{action-line}
\end{eqnarray}
Here we used a prime to denote derivation by $r$, and $P[C_4]$ is the 
pullback of the four-form to the world-volume. The tension of the 
D3-brane,  $T_{\rm D3}$, is given by
\begin{equation}
T_{\rm D3}=\frac{1}{(2\pi)^3l_s^4g_s}
=\frac{N}{2\pi^2L^4}\,.
\end{equation}

Since the world-volume gauge field $A_1$ does not appear explicitly
in the action, its conjugate momentum $i\Pi$ is conserved (in the 
Euclidean theory the electric field is imaginary, so with this extra $i$ 
we will get a real quantity). It  is
\begin{equation}
\Pi=-i\frac{4N}{\lambda}\frac{2\pi F_{1r}r^2}
{\sqrt{1+y'^2+4\pi^2F_{1r}^2y^4/\lambda}}\,.
\label{Pi-line}
\end{equation}
In this definition we integrated the momentum over the $S^2$, which 
gave the conserved charge corresponding to the fundamental string 
density. So $\Pi$ will be an integer, $k$, which corresponds to the 
number of coincident Wilson loops.

Motivated by the spike solution in flat space 
\cite{Callan}\footnote{
In flat space the transverse coordinate behaved like $X^9\sim 1/r$. 
This coordinate $X^9$ is replaced here by $1/y$, hence the linear 
ansatz.}, 
we consider the linear 
ansatz $y=r/\kappa$. Plugging this into the equations of motion we 
find that it solves them for the constant 
$\kappa=k\sqrt\lambda/4N$. This gives the electric field
\begin{equation}
F_{1r}=i\frac{k\lambda}{8\pi N r^2}\,.
\label{F-line}
\end{equation}
This solution is in fact a limit of one described in \cite{Gauntlett}, 
where they consider a D3-brane probe in the background generated by 
other D3-branes. Ours is just the near horizon limit of their solution.

As stated, the solution is a hypersurface in $AdS_5$, and the induced 
metric on the brane is given by
\begin{equation}
ds^2=\frac{L^2\kappa^2}{r^2}
\left((1+\kappa^{-2})dr^2+dt^2\right)
+L^2\kappa^2\left(d\theta^2+\sin^2\theta d\phi^2\right)\,.
\end{equation}
Thus the geometry has the product structure\footnote{
D-brane solutions with $AdS\times S$ induced geometry have appeared 
before in the literature, see e.g. \cite{bachas,Skenderis:2002vf}; 
an important difference is that 
in our case the entire brane, including the $S^2$, is embedded in $AdS_5$, 
while in those examples the sphere part is inside the sphere of 
target space.}
$AdS_2\times S^2$. The curvature radius of the $AdS_2$ factor is 
$L\sqrt{1+\kappa^2}$ and of the $S^2$ is $L\kappa$.
Having this product structure means that the sphere never shrinks, 
even as we approach the boundary of $AdS_5$. But in the dual CFT it 
corresponds to a point, and not 
to a finite size sphere due to the infinite rescaling of the metric near 
the boundary of $AdS$ \cite{Witten}.

Calculating the action for this solution we find that the WZ term 
exactly cancels the DBI part, giving $S=0$. This is 
the expected final answer, but the calculation is not complete. Thus 
far we only considered the bulk action, one should add to this appropriate 
boundary terms, to which we will turn now.

\subsection {Boundary terms}

The D3-brane solution we found extends all the way to the boundary 
of $AdS_5$ and ends there along a one-dimensional curve. This opens up 
the possibility of adding boundary terms to the action. These boundary
terms don't change the equations of motion, so the solution is still the same,
but the value of the action when evaluated at this solution will in general 
depend on the boundary terms.

When calculating the Wilson loop using string surfaces the bulk action 
is divergent, but this term is fixed by a boundary term \cite{hirosi}. 
Let us recall the argument, since we will have to apply it for the 
case at hand.

The string used to describe the Wilson loop has to satisfy complementary 
boundary conditions with respect to a free string ending on a D3-brane. 
While the latter has to satisfy Dirichlet conditions for six 
directions and Neumann conditions for the other four, the string 
describing the Wilson loop has to satisfy six Neumann conditions and 
four Dirichlet. An easy way to convince oneself of this fact is to 
consider a Wilson loop on a D9-brane, which has to follow a curve 
in ten-dimensions, hence ten Dirichlet conditions, after six T-dualities 
one gets the D3-brane and the boundary conditions stated above.

The Dirichlet boundary conditions are on the four directions 
parallel to the boundary of $AdS$, and the six Neumann ones combine 
the radial coordinate of $AdS$ and the $S^5$ coordinates. The 
Nambu-Goto action for the string (as well as the Polyakov action) 
is a functional of the coordinates, which is the appropriate action 
assuming we have Dirichlet boundary conditions. So we have to 
add boundary terms that change the boundary conditions.

Since all the Wilson loops we discuss have no dependence on the 
$S^5$, the only coordinate we have to replace with its momentum is 
the radial coordinate $y$. We therefore define $p_y$ as the momentum 
conjugate to it
\begin{equation}
p_y= \frac{\delta S}{\delta \partial _n y}\,,
\end{equation}
where $\partial_n$ is the normal derivative to the boundary.

The new action including the term that changes the boundary conditions 
is
\begin{equation}
\tilde S=S-y_0\int d\tau\, p_y\,,
\end{equation}
where the integral is over the boundary at a cutoff $y=y_0$. 
The original action $S$ is a functional of $y$ and $\partial y$. 
Applying the standard variational technique to it we find
\begin{equation}
\delta S=\int d^2\sigma\left[\frac{\delta S}{\delta y}\delta y
+\frac{\delta S}{\delta \partial y}\delta \partial y\right]
=\oint d\tau\, p_y\delta y\,,
\end{equation}
where the bulk part part vanishes due to the equations of motion. The 
boundary term clearly indicates that it is a functional of $y$. Including the 
boundary term, the variation of the new action is
\begin{equation}
\delta\tilde S=-\oint d\tau\, y \delta p_y\,,
\end{equation}
and it's indeed a functional of $p_y$, as advertised.

When calculating the action for the fundamental string this boundary 
term cancels the divergence in the area. In our calculation, using 
D3-branes the action was finite (zero), so we do not need to cancel 
a divergence, but the logic that applied to the string still holds, and 
we should apply the same procedure here. The DBI action is a 
functional of the coordinates, and in particular of $y$ and $y'$, so 
we have to add the same kind of boundary term as for the string.

Using our action (\ref{action-line}) we find that the momentum 
conjugate to $y$ (integrated over the sphere) is
\begin{equation}
p_y=\frac{2N}{\pi}\frac{r^2y'}{y^4
\sqrt{1+y'^2+(2\pi \alpha' F_{1r})^2 \frac{y^4}{L^4}}}\,.
\end{equation}
Using the equations of motion we get the boundary term inserted at 
a cutoff $y_0$
\begin{equation}
-\int dx^1\, y_0 p_y
=-\frac{2N}{\pi} \frac{X^1\kappa}{y_0}\,.
\label{DBI-boundary}
\end{equation}
Note that for the bulk part of the DBI action diverged like 
$\kappa^3/y_0$, so for small $\kappa$ the boundary term 
is much larger than the bulk contribution. It is actually the same as 
for the fundamental string. There it exactly canceled the bulk term, 
since the world-sheet was perpendicular to the boundary, i.e. in the 
$y$ direction. The D3-brane extends also in the $r$ direction and 
has non-zero momentum in that direction, that is why $p_y$ is not 
equal to the bulk action.

Since we still expect the full action to vanish, there must be another 
boundary term, which is the Legendre transform of the other 
variable, 
the gauge field. The action (\ref{action-line}) is a functional of the 
gauge field, but the Wilson loop observable defines the number $k$, 
which is the dimension of the representation of the loop, or the 
wrapping number, for a multiply wound loop. The momentum 
conjugate to the the gauge field, $\Pi$, calculated in (\ref{Pi-line}) 
is precisely equal to $k$, therefore it's the correct variable to use, 
instead of the gauge field. Thus we find another boundary 
contribution to the action, which we write as the integral over the 
total derivative
\begin{equation}
-\int dx^1\, i\Pi A_1
=-\int dx^1\,dr\, i\Pi F_{1r}
=\frac{2N}{\pi}\frac{X^1\kappa}{y_0}\,.
\end{equation}

The sum of the two Legendre transforms - of the coordinate $y$ 
and of the gauge field - add up to zero. So on-shell there is no boundary 
contribution to the action and we are left with a total action $S=0$, or 
$\vev{W}=1$, as expected from supersymmety.

There is another subtlety associated with the boundary. The Wess-Zumino 
part is not well defined on a manifold with boundary. It should be 
the integral of the five-form flux surrounded by the brane
\begin{equation}
S_{\rm WZ}=-T_{\rm D3}\int _{M_5} F_5\,,
\end{equation}
For a D3-brane without 
boundaries this is well defined (up to replacing the inside and outside), 
but it's not clear what $M_5$ should be for a D3-brane with boundary. 

We defined the action in terms of the pullback of the 4-form potential
\begin{equation}
S_{\rm WZ}=-T_{\rm D3}\int_{M_4} P[C_4]\,.
\end{equation}
Under gauge transformations $\delta C_4=d\Lambda$ (since the 
NS 2-form vanishes). So the variation of the action is
\begin{equation}
\delta S_{\rm WZ}=-T_{\rm D3}\int_{M_4} P[\delta C_4]
=-T_{\rm D3}\int_{\partial M_4} P[\Lambda]\,.
\end{equation}
We used a very natural form of $C_4$ (\ref{C4}) 
possessing the relevant symmetries 
of our problem and no singularities along the Wilson loop. Other choices 
may lead to different answers (see the discussion in section 
\ref{conf-trans-sec}). One may fix this ambiguity by hand---adding 
a three-form on the boundary (equal to zero in our gauge) and 
imposing that it transforms by $P[\Lambda]$ under 
gauge transformations. It would be good to get a better understanding 
of this issue.

\subsection{Supersymmetry}

The infinite straight Wilson loop preserves half of the supersymmetries 
of the theory, as does the string solution in $AdS$. Here we will show 
that the D3-brane preserves the same supersymmetries.

In order to check supersymmetry we switch to Lorentzian signature and 
define the vielbeins
\begin{equation}
e_y^{\bar y}=\frac{L}{y}\,,\qquad
e_t^{\bar t}=\frac{L}{y}\,,\qquad
e_r^{\bar r}=\frac{L}{y}\,,\qquad
e_\theta^{\bar\theta}=\frac{Lr}{y}\,,\qquad
e_\phi^{\bar\phi}=\frac{Lr\sin\theta}{y}\,.
\end{equation}

We use $\Gamma_a$ as constant gamma matrices and define 
$\gamma_\mu=e_\mu^a\Gamma_a$. Using two constant spinors of 
positive and negative chirality $\epsilon_0^\pm$ that satisfy also 
$i\Gamma_{\bar t\bar r\bar\theta\bar\phi}\epsilon _0^{\pm}
=\pm \epsilon_0^{\pm}$. 
and the matrix
\begin{equation}
M=\exp\left(\frac{\theta}{2}\Gamma_{\bar r\bar\theta}\right)
\exp\left(\frac{\phi}{2}\Gamma_{\bar\theta\bar\phi}\right)\,,
\end{equation}
the Killing spinors of $AdS_5$ are written as\footnote{
Since our solution has no dependence on the $S^5$ part of the geometry, 
we do not need the form of the Killing spinors on the full 
$AdS_5\times S^5$. To account for that, one simply has to multiply 
the constant spinors $\epsilon_0^\pm$ with a function of the $S^5$ 
coordinates and gamma matrices.}
(see for example \cite{Skenderis:2002vf})
\begin{equation}
\epsilon=y^{-1/2}M\epsilon_0^-
+\left(y^{1/2}\Gamma_{\bar y}
+y^{-1/2}(r\Gamma_{\bar r}+t\Gamma_{\bar t})\right)M\epsilon_0^+\,.
\end{equation}
They satisfy the equation $D_\mu\epsilon
=\frac{i}{2L}\Gamma_{\bar t\bar r\bar\theta\bar\phi\bar y}
\gamma_\mu\epsilon$.

The supersymmetries preserved by the D3-brane are generated by the 
Killing spi\-nors that also satisfy $\Gamma\epsilon=\epsilon$ 
where $\Gamma$ is the projector associated with the D3-brane. In 
our case it is given by
\begin{equation}
\Gamma =\frac{1}{\sqrt{-\det(g+2\pi\alpha' F)}}
\left[(y'\gamma_{ty\theta\phi}
+\gamma_{tr\theta\phi})I
-2\pi\alpha'F_{tr}\gamma_{\theta\phi}KI\right]\,,
\end{equation}
with $K$ acts on spinors by complex conjugation and $I$ multiplies them 
by $-i$. For our solution the square root in the denominator is equal 
to $L^4r^2\sin\theta/y^4$ 
while $y'=1/\kappa$ and in the Lorentzian theory 
$2\pi\alpha'F_{tr}=L^2\kappa/r^2$.

Naively it would seem like the equation $\Gamma\epsilon=\epsilon$ 
would impose two conditions on the spinors, so only $1/4$ of the 
supersymmetries would be preserved. This is in fact what happens 
in the flat space case of Callan and Maldacena \cite{Callan}. But 
since $AdS$ is the background created by D3-branes, the projector 
associated with the D3-brane does not break any of the supersymmetries.

To see that we rewrite $\Gamma$ as
\begin{equation}
\Gamma=\left[1-\kappa^{-1}\Gamma_{\bar r}
\left(\Gamma_{\bar y}-\Gamma_{\bar t}K\right)\right]
\Gamma_{\bar t\bar r\bar\theta\bar\phi}I\,,
\end{equation}
and the simplification arises since $\epsilon_0^\pm$ are eigenstates 
of $\Gamma_{\bar t\bar r\bar\theta\bar\phi}I$. A bit of algebra 
gives
\begin{equation}
\Gamma\epsilon-\epsilon
=-\kappa^{-1} y^{-1/2}\Gamma_{\bar r}M
(\Gamma_{\bar y}-\Gamma_{\bar t}K)\epsilon_0^-
+\kappa^{-1} y^{-1/2}\Gamma_{\bar r}
\left(t\Gamma_{\bar t}-r\Gamma_{\bar r}-y\Gamma_{\bar y}\right)
M\left(\Gamma_{\bar y}+\Gamma_{\bar t}K\right)\epsilon_0^+\,.
\end{equation}
Thus the system will be invariant under supersymmetries generated by 
$\epsilon$ made up of $\epsilon_0^\pm$ subject to the constraints
\begin{equation}
\Gamma_{\bar y}\epsilon_0^-=\Gamma_{\bar t}{\epsilon_0^-}^*
\qquad
\Gamma_{\bar y}\epsilon_0^+=-\Gamma_{\bar t}{\epsilon_0^+}^*\,,
\end{equation}
where ${\epsilon_0^\pm}^*$ is the complex conjugate of 
$\epsilon_0^\pm$.

\section{Circular loop}

After proving the feasibility of using a D3-brane to calculate a Wilson 
loop in the case of the straight line, we turn now to the more 
interesting case of the circular Wilson loop.

\subsection{Bulk calculation}

Let's start with the coordinate system for $AdS_5$
\begin{equation}
ds^2=\frac{L^2}{y^2}\left(dy^2+dr_1^2
+r_1^2d\psi^2+dr_2^2+r_2^2d\phi^2\right)\,,
\label{r1r2}
\end{equation}
where $r_1$ is the radial coordinate in the $x^1,x^2$ plane and $r_2$ is 
the radial coordinate in the $x^3,x^4$ plane. We place the Wilson loop
at $r_1=R$, and $r_2=0$. We want to find a D3-brane solution of the DBI
action, pinching to this circle as $y\rightarrow 0$. To find the 
solution, it turns out to be more convenient
to change to the coordinates $\rho$, $\theta$, and $\eta$ defined
by\footnote{These coordinates have some resemblance to the ones used to 
describe black rings, see e.g. \cite{Emparan:2001wn}.}
\begin{equation}
r_1=\frac{R\cos\eta}{\cosh\rho-\sinh\rho\cos\theta}\,,
\quad
r_2=\frac{R\sinh\rho\sin\theta}{\cosh\rho-\sinh\rho\cos\theta}\,,
\quad
y=\frac{R\sin\eta}{\cosh\rho-\sinh\rho\cos\theta}\,.
\label{coords}
\end{equation}
With these coordinates the metric of $AdS_5$ is given by
\begin{equation}
ds^2=\frac{L^2}{\sin^2\eta}
\left(d\eta^2+\cos^2\eta d\psi^2+d\rho^2+\sinh^2\rho
\left(d\theta^2+\sin^2\theta d\phi^2\right)\right)\,.
\label{circle-metric}
\end{equation}
These coordinates cover the space once if they take the ranges 
$\rho\in[0,\infty)$, $\theta\in[0,\pi]$ and $\eta\in[0,\pi/2]$. 
The boundary of space $y=0$ is mapped to $\eta=0$ as well as 
$\rho\to\infty$. The circle on the boundary is located 
at $\eta=\rho=0$.

The string solution describing the circular Wilson loop is given by 
$r_1^2+y^2=R^2$, or in the new coordinate system by $\rho=0$.
The bulk action is then
\begin{equation}
S_{\rm bulk}=\frac{\sqrt\lambda}{2\pi}\int d\eta\,d\psi\,
\frac{\cos\eta}{\sin^2\eta}
=\sqrt\lambda\left(\frac{1}{\sin\eta_0}-1\right)\,,
\end{equation}
with $\eta_0$ a cutoff on $\eta$. 
The divergence is removed by the boundary term and we are left with 
the answer
\begin{equation}
\vev{W_{\rm circle, string}}=\exp\sqrt\lambda\,.
\end{equation}

We wish now to find the appropriate D3-brane that ends along the 
circle. Again it will be given by a single equation, and the symmetries 
guarantee that it is of the form $\eta=\eta(\rho)$. So we may take 
$\psi$, $\rho$, $\theta$ and $\phi$ as the world-volume coordinates. 
There is a single scalar field $\eta$ and a single component of the 
electromagnetic field $F_{\psi\rho}(\rho)$. We take the four-form 
potential, $C_4$, to be the same as for the straight line.
In the $(r_1, r_2)$ coordinates this is just
\begin{equation}
C_4=L^4\frac{r_1r_2}{y^4}
dr_1\wedge d\psi\wedge dr_2\wedge d\phi\,.
\end{equation}
and in the new coordinates it is
\begin{eqnarray}
C_4 &=&
L^4\frac{\cos^2\eta\sin\theta\sinh^2\rho}
{\sin^4\eta}
d\rho\wedge d\psi\wedge d\theta\wedge d\phi
\nonumber\\&&
+L^4\frac{\cos\eta\sin\theta\sinh^2\rho
(\sinh\rho-\cosh\rho\cos\theta)}
{\sin^3\eta(\cosh\rho-\sinh\rho\cos\theta)}
d\eta\wedge d\psi\wedge d\theta\wedge d\phi
\nonumber\\&&
-L^4\frac{\cos\eta\sin^2\theta\sinh\rho}
{\sin^3\eta(\cosh\rho-\sinh\rho\cos\theta)}
d\eta\wedge d\psi\wedge d\rho\wedge d\phi
\,.
\end{eqnarray}

Using this the DBI part of the action is
\begin{eqnarray}
S_{\rm DBI}&=&T_{\rm D3}
\int e^{-\Phi}\sqrt{\det(g+2\pi\alpha' F)}
\\
&=&2N\int d\rho\,d\theta\,
\frac{\sin\theta\sinh^2\rho}{\sin^4\eta}
\sqrt{\cos^2\eta(1+\eta'^2)+
(2\pi \alpha')^2\frac{\sin^4\eta}{L^4}F_{\psi\rho}^2}\,.
\nonumber
\label{DBI-circle}
\end{eqnarray}
The WZ part is
\begin{eqnarray}
S_{\rm WZ}&=&-T_{\rm D3}\int P[C_4]
\label{WZ-circle}
\\
&=&-2N\int d\rho\,d\theta\,
\frac{\cos\eta\sin\theta\sinh^2\rho}{\sin^4\eta}\left(
\cos\eta+\eta'\sin\eta\,\frac{\sinh\rho-\cosh\rho\cos\theta}
{\cosh\rho-\sinh\rho\cos\theta}\right).
\nonumber
\end{eqnarray}

Again $\Pi$, the momentum conjugate to the gauge field, is conserved, 
and is equal to the fundamental string charge, or the number of 
coincident Wilson loops. Now the ansatz 
$\sin\eta=\kappa^{-1}\sinh\rho$ solves the equations of motion 
if\footnote{Again we defined $\Pi$ as $-i$ times the conjugate to 
$F$, to make it real. Also it is defined integrated over $\theta$ and 
$\phi$, but not over $\psi$, so it corresponds to the effective 
fundamental string density.}
\begin{equation}
k=\Pi=\frac{4N\kappa}{\sqrt\lambda}\,,
\qquad\hbox{which gives}\qquad
F_{\psi\rho}=\frac{ik\lambda}{8\pi N\sinh^2\rho}\,,
\end{equation}
It should be now obvious why we chose this coordinate system. For 
one it preserves the symmetry of the problem, as there is no $\theta$ 
dependence. But more than that, near the boundary, where $\eta$ is 
small we see that the linear approximation to the solution is the same 
as for the straight line, with the replacements $\rho\to r$ and 
$\eta\to y$. This clearly will always be true, since in the UV all smooth 
loops look like the straight line.

The induced metric on the D3-brane is given by 
\begin{equation}
ds^2=\frac{L^2}{\sin^2\eta}\left(
\frac{1+\kappa^2}{1+\kappa^2\sin^2 \eta}d\eta^2
+\cos^2\eta d\psi^2\right)
+L^2\kappa^2\left(d\theta^2+\sin^2\theta d\phi^2\right)\,,
\end{equation}
which again is the metric of $AdS_2\times S^2$ (to see that one 
can switch to the coordinate $\zeta$ defined by 
$\cot^2\eta=(1+\kappa^2)\sinh^2\zeta$). As before the radius of 
the $AdS_2$ factor is $L\sqrt{1+\kappa^2}$ and of the sphere 
$L\kappa$. The main difference is that here we find the global 
structure of $AdS_2$ is the Poincar\'e disc, while before it was 
the upper half plane. This is the same kind of difference found between 
the string solutions describing those two Wilson loops.
The D3-brane hypersurface is depicted in Fig. 1. in the $r_1$, $r_2$ 
and $y$ coordinate system.

Like the straight line, this solution also preserves half the 
supersymmetries.

\figin{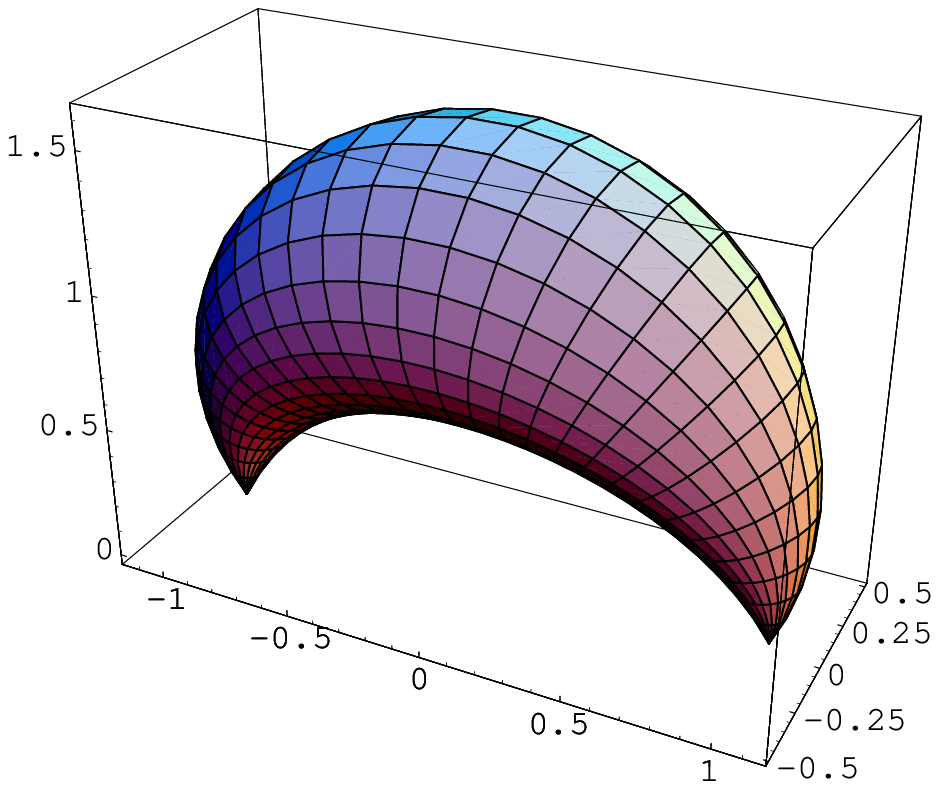}{A depiction of the D3-brane solution describing the 
circular Wilson loop of radius $R=1$. 
The horizontal plane in the figure is the $r_1,r_2$ 
plane and the vertical direction is $y$, the two angular directions 
$\psi$ and $\phi$ are suppressed. The surface pinches off at the 
boundary of $AdS$ (the bottom of the picture) at $r_2=0$ and $r_1=1$ 
(also $r_1=-1$), and expands away from it.}

Both the DBI and WZ parts of the action suffer from a 
divergence near the boundary, but plugging the solution into the 
action we find the combination to be perfectly finite
\begin{eqnarray}
S_{\rm DBI+WZ}&=&2N\kappa^2\int d\rho\, d\theta\,
\frac{\sin\theta\cos\theta}
{\sinh\rho(\cosh\rho-\sinh\rho\cos\theta)}
\nonumber\\
&=&2N\kappa ^2
\left [\coth\rho-\frac{\rho}{\sinh^2\rho}
\right]_{\rho=0}^{\sinh \rho =\kappa}\,.
\label{bulk-circle}
\end{eqnarray}
There is no contribution from the lower limit of the integral, near the 
boundary. So the bulk action is the above expression evaluated at 
$\sinh\rho=\kappa$. Before studying it further we turn to the 
boundary terms.

\subsection{Boundary terms}

As in the case of the straight line we have to complement the bulk 
calculation with boundary terms. The first of these is the Legendre 
transform of the radial coordinate in $AdS$. In our coordinate system 
(\ref{circle-metric}) the boundary is given by $\eta\to0$ (also 
$\rho\to\infty$, but that regime is far from our D3-brane). We 
are more used to the radial coordinate $y$ of (\ref{r1r2}), 
but the two are proportional to each other near the boundary, up 
to $O(y^3)$ corrections. So the prescription for the Wilson loop 
involves the Legendre transform of $\eta$.

The momentum conjugate to $\eta$ is\footnote{Here it is defined 
as not integrated over the $S^2$, since it has nontrivial $\theta$ 
dependence.}
\begin{equation}
p_\eta=\frac{\delta L}{\delta \eta'}
=\frac{N}{2\pi^2}
\frac{\sin\theta\sinh^2\rho\cos^2\eta}{\sin^4\eta}
\left[\eta'-\tan\eta\,\frac{\sinh\rho-\cosh\rho\cos\theta}
{\cosh\rho-\sinh\rho\cos\theta}\right],
\end{equation}
where the first term comes from the DBI part and the second from 
the WZ piece. The resulting boundary term is
\begin{equation}
-\eta_0\int p_\eta
=-4N\frac{\kappa}{\eta_0}+O(\eta_0)\,.
\end{equation}
Here the WZ part contributed only terms that vanish in the 
$\eta_0\to0$ limit, and the result is the same as for the straight line 
(except that the length of the line $X^1$ is replaced by $2\pi$).

Next we perform the Legendre transform on the gauge field, replacing 
it with the conjugate momentum $\Pi=k$. As in the straight line case 
we do that by adding the total derivative
\begin{equation}
-\int d\rho\,d\psi\,i\Pi F_{\psi\rho}
=-4N\kappa^2\coth\rho
\Bigg|_{\sinh\rho=\kappa\sin\eta_0}^{\sinh\rho=\kappa}\,.
\end{equation}
The contribution from the lower limit is 
equal to $4N\kappa/\eta_0$ and exactly cancels the divergence from 
the other boundary term. The contribution from the upper limit combines 
with the bulk term (\ref{bulk-circle}) to give
\begin{equation}
S_{\rm circle}=
-2N\left [\kappa\sqrt{1+\kappa^2}
+\sinh^{-1}\kappa\right].
\label{final}
\end{equation}

As before there should also be a boundary term to make the Wess-Zumino 
part of the action gauge invariant. Since we did not find a compelling 
expression for this term, we leave it open.

\subsection{Analysis of the result}

The resulting expression for the action of the D3-brane 
(\ref{final}) is rather 
complicated, let us expand it in the regime we are most familiar with, 
where $\lambda\ll N^2$ and $k$ is small, or small $\kappa$. 

This gives
\begin{equation}
S_{\rm circle}=
-4N\kappa
-\frac{2N\kappa^3}{3}
+\frac{N\kappa^5}{10}
+O(\kappa^{7})
=-k\sqrt\lambda
-\frac{k^3\lambda^{3/2}}{96N^2}
+\frac{k^5\lambda^{5/2}}{10240N^4}
+O\left(\frac{\lambda^{7/2}}{N^6}\right).
\label{expansion}
\end{equation}
The first term in the final expression is the same as the action of $k$ 
coincident strings (\ref{circle-string}), but the full result includes 
an infinite series of corrections in $1/N^2$. Can those terms be 
explained in terms of the fundamental strings?

The explanation to these terms was given in fact in 
\cite{Drukker:2000rr}, so let us review it. We want to examine 
the string loop contributions to the Wilson loop as calculated using 
fundamental strings in $AdS$. At large $\lambda$ we 
would be instructed to find classical minimal surfaces of higher genus 
ending on the curve on the boundary. Such smooth solutions will 
not exist, instead we should consider the original solution with 
degenerate handles attached to it.

So at order $g_s^{2p}$ we should consider $p$ indistinguishable degenerate 
handles ending on our surface. They will have the same action as the 
leading planar result, but with a different prefactor. The string coupling 
gives the obvious factor $g_s^{2p}\sim (\lambda/N)^{2p}$. In addition 
we should worry about the measure of integration. 
A generic open string with one boundary and $p$ handles will have 
$6p-3$ real moduli. In the large $\lambda$ limit we have to consider only 
degenerate handles, which imposes two constraints per handle (so each 
handle is left with four real moduli, the locations of its two ends). Each of 
those constraints (in addition to the overall three) introduce a delta 
function into the integration measure that will give a factor of 
the inverse effective length-scale of the problem, i.e.
$\lambda^{-1/4}$. Together with the combinatorial factor we expect 
therefore at order $2p$ a result proportional to
\begin{equation}
\frac{1}{p!}
\frac{g_s^2}{\lambda^{(2p+3)/4}}
\sim
\frac{1}{p!}
\frac{\lambda^{(6p-3)/4}}{N^{2p}}\,.
\end{equation}
Those corrections will all exponentiate to give a term proportional to 
$\lambda^{3/2}/N^2$ in the expectation value of the Wilson loop. 
If there are $k$ coincident string surfaces our calculation shows 
that the result will scale with $k^3$, but we don't have a good 
heuristic argument for this scaling.

Getting this term assumed that the handles are all independent. There 
will be also contributions when two (or more) of those handles collide, 
which are even more degenerate surfaces. Those surfaces will have 
higher genus, and smaller measure, resulting in terms like 
$\lambda^{5/2}/N^4$ and so on.

It is amusing to look separately at the bulk and boundary contributions. 
The bulk contribution alone gives
\begin{equation}
S_{\rm circle, bulk}= \frac{k^3\lambda^{3/2}}{48N^2}
+O\left(\frac{k^5\lambda^{5/2}}{N^4}\right)\,,
\end{equation}
so it does not include the term linear in $k$. To get those correctly it 
was crucial to include the boundary terms that made the action a functional 
of the correct variables, the momenta $p_\eta$ and $\Pi$.

We have proposed a different prescription for calculating Wilson loops, 
by using D3-branes rather than fundamental strings. Are both calculations 
equally good, or is there a reason to prefer one over the other?

This issue was addressed in the flat-space case \cite{Callan} and 
two criteria were found for the validity of the D-brane calculation. The 
first requirement was that the fields on the brane will be slowly varying, 
and the other is that the system does not back-react on the geometry.

Looking at the D3-brane world-volume, it has the product structure 
$AdS_2\times S^2$. The radius of curvature of the $AdS_2$ factor is 
$L\sqrt{1+\kappa^2}$, so it never becomes small. The radius of 
curvature of the $S^2$ part is $L\kappa $. The 
calculation cannot be trusted unless this radius is larger than $l_s$,
which translates into requiring $\kappa\gg\lambda ^{-1/4}$. To 
prevent the system to back-react on the geometry we have to impose 
$kg_s^2\ll1$. In terms of our variables this translates to 
$\kappa\ll1/(g_s\sqrt\lambda)$. Note that since we always assume 
$\lambda\gg1$, the range of validity includes the regime of small 
$\kappa$, where the first term, $-4N\kappa=-k\sqrt\lambda$ 
dominates.

In some ways the D3-brane solution in $AdS$ is better than in flat space 
\cite{Callan}. In flat space the D3-brane world-volume gets highly curved 
near the source of the electric field, and the field strength and its 
derivatives diverge. The induced metric on our solution is homogeneous, 
so our requirement will lead to small curvature on the entire brane, not 
only in some asymptotic part.

The specific cases studied in this paper are supersymmetric, and the 
calculation seems to work beyond the expected range of validity. As we 
will see in the next subsection, the result we obtained matches with a
matrix model computation, assuming just $\lambda \gg 1$ with no 
other restrictions on $\kappa$.

Let us comment about the magnetic case (studied below). The requirement 
of no back-reaction is now $k\tilde g_s\ll1$, which we wrote in terms of 
the dual couplings $\tilde g_s=1/g_s$. In the magnetic 
case $\kappa$ is defined as $\kappa=\pi k/\sqrt{\tilde\lambda}$ 
(where $\tilde\lambda=16\pi^2N^2/\lambda$) which leads to the 
same requirement as in the electric case $\kappa\ll1/(g_s\sqrt\lambda)$.
The condition on the radius of the sphere will again be 
$\kappa\gg\tilde\lambda^{-1/4}$, so with those definitions of 
$\kappa$ we find the same range of validity for the electric and magnetic 
cases.

\subsection{Comparison with the matrix model}

As already mentioned before, the circular Wilson loop has very nice 
properties when calculated in perturbation theory. The combined 
propagator of the gauge field and scalar is a constant, which reduces 
the calculation of all rainbow/ladder diagrams to a zero-dimensional 
matrix model \cite{erickson}, which can be written explicitly as an 
integral over all $N\times N$ Hermitian matrices $M$
\begin{equation}
\vev{W_{\rm ladders}}=\vev{\frac{1}{N} \Tr\exp M}
=\frac{1}{Z}\int {\cal D}M
\frac{1}{N} \Tr e^M e^{-\frac{2N}{\lambda} \Tr M^2}
\end{equation}
The leading behavior at large $N$, expressed in terms of the Bessel 
function $I_1$, is easily found using Wigner's 
semi-circle law and is 
\begin{equation}
\vev{W_{\rm ladders}}
\sim \int_{-1}^1
dx\, \sqrt{1-x^2}\,e^{x\sqrt{\lambda}}
=\frac{2}{\sqrt\lambda}I_1\left(\sqrt\lambda\right)
\sim e^{\sqrt\lambda}\,,
\end{equation}
which is indeed the leading behavior of the circular Wilson loop as 
calculated by a string in $AdS$.

One can do better and solve this matrix model exactly applying several 
different 
techniques. In \cite{Drukker:2000rr} this was done using orthogonal 
polynomials to give the full result at finite $N$, then the result was 
rearranged in a $1/N$ expansion. Using those expressions the leading 
$\sqrt\lambda$ term was reproduced as well as an infinite series 
of corrections. It was noticed there that some of the terms in this 
series exponentiated to give $\exp[\lambda^{3/2}/(96 N^2)]$, exactly 
as in (\ref{expansion}). From equation (B.7) of that paper the 
next term, $\lambda^{5/2}/(10240N^4)$, can also be extracted with 
the correct coefficient.

To check our result more closely let us recall the result for the 
matrix model at finite $N$. It was given in terms of a Laguerre 
polynomial\footnote{
$L_n^k(x)=1/n!\exp[x]x^{-k}(d/dx)^n(\exp[-x]x^{n+k})$}
as
\begin{equation}
\vev{W_{\rm ladders}}
=\vev{\frac{1}{N}\Tr\exp M}
=\frac{1}{N}L_{N-1}^1(-4N\kappa^2)\exp[2N\kappa^2]\,,
\label{laguerre}
\end{equation}
The multiply wound Wilson loop is given by the expectation 
value of the matrix model operator $\Tr\exp kM$, which amounts 
to the replacement $\lambda\to k^2\lambda$. That allowed us 
to express $\lambda$ in the matrix model result in terms of 
$N$ and $\kappa$.

One could consider other operators, like $(\Tr \exp M)^2$, which 
would correspond to two overlapping Wilson loops. While the planar 
result will always scale with the total multiplicity $\exp k\sqrt\lambda$, 
the exact expressions are more complicated \cite{Drukker:2000rr}, 
and one should not expect the same answer as the multiply wrapped 
loop beyond the planar limit.

Going back to the case of the single trace operator, it is expressed in terms 
of a Laguerre polynomial which satisfies the differential equation
\begin{equation}
xL_n^k(x)''+(k+1-x)L_n^k(x)'+nL_n^k(x)=0\,.
\end{equation}
This leads to the following differential equation for the expectation 
value of the Wilson loop
\begin{equation}
\left[\kappa\partial_\kappa^2+3\partial_\kappa
-16N^2 \kappa(1+\kappa^2)\right]
\vev{W_{\rm ladders}}=0\,.
\end{equation}
Next we rewrite our observable as the exponent of an effective 
action $\vev{W_{\rm ladders}}=\exp[-N\cF]$ and derive the equation 
for $\cF(\kappa)$
\begin{equation}
(\cF')^2
-\frac{1}{N\kappa}\left(\kappa \cF''+3\cF'\right)-
16(1+\kappa^2)=0\,.
\end{equation}
Since $N\kappa \sim k\sqrt{\lambda}$ and we are in the regime where 
$\lambda \gg 1$, we will neglect the terms subleading in $N$ and 
$N\kappa$, to find
\begin{equation}
\frac{d\cF}{d\kappa}=\pm4\sqrt{1+\kappa^2}\,.
\end{equation}
Finally we integrate this to find
\begin{equation}
\cF=\cF_0\pm2\left[\kappa\sqrt{1+\kappa^2}+\sinh^{-1}\kappa\right]\,.
\end{equation}
Fixing the boundary condition $\cF_0=0$ and the sign $-$ from the 
explicit results of the matrix model stated above, we get full 
agreement with the D3-brane calculation (\ref{final})
\begin{equation}
S=N\cF\,.
\end{equation}

Notice that to match both computations, we only had to require 
$N\kappa \gg 1$, or equivalently $\lambda \gg 1$. So the match works
even for $k=1$.

\subsection{Conformal transformation}
\label{conf-trans-sec}

While we presented the calculation of the circular Wilson loop as 
independent from the straight line, the two are actually intimately 
connected. After all, the straight line and the circle are related by
a conformal transformation, and that conformal transformations in the
boundary of $AdS$ extend to isometries of the full $AdS$ space. 
Therefore, one may obtain the D3-worldvolume associated to the
circular Wilson loop by the (extension of the) conformal transformation
that takes the line to the circle\footnote{Actually, this is how we 
originally found the solution}. It is illuminating to carry out this exercise.

The special conformal transformations generated by a vector $c^\alpha$ 
on the boundary is extended to the isometry of $AdS$
\begin{eqnarray}
x^\alpha&=&
\frac{\tilde x^\alpha+c^\alpha((\tilde x)^2+\tilde y^2)}
{1+2c\cdot\tilde x +(c)^2((\tilde x)^2+\tilde y^2)}\,,
\nonumber\\
y&=&\frac{\tilde y}
{1+2c\cdot\tilde x +(c)^2((\tilde x)^2+\tilde y^2)}\,.
\end{eqnarray}
The inverse is given by the same equations with $c^\alpha\to -c^\alpha$.

Starting with the straight line we first have to shift it away from the origin 
in the $x^2$ direction to $(x^1,1/2,0,0)$, then using the above transformation 
(with $c=(0,-1,0,0)$), and finally rescale all the coordinates by a factor 
$R$, we find that the hypersurface defined by the equation $r=\kappa y$
is transformed to
\begin{equation}
(r_1^2+r_2^2+y^2-R^2)^2+4R^2r_2^2=4\kappa^2R^2y^2\,.
\end{equation}
Indeed, at the boundary of $AdS$, $y=0$, this hypersurface reduces to the 
circle with $r_1=R$, $r_2=0$. Writing this equation using the coordinates 
in (\ref{circle-metric}) gives the solution found above: 
$\sin\eta=\kappa^{-1}\sinh\rho$.

Another way to find these hyper-surfaces is to notice that they are the 
orbits of the $SL(2,\bR)\times SU(2)$ subgroup of the four dimensional 
conformal group preserved by the circular loop \cite{Bianchi:2002gz}. 
Clearly this symmetry acts naturally on the $AdS_2\times S^2$ surfaces 
we found.

We can also apply the conformal transformation to the four-form potential 
(this is most easily done in Cartesian coordinates), obtaining
\begin{equation}
C_4'=\frac{L^4r_2}{y^4}(r_1 dr_1+ydy)\wedge d\psi \wedge dr_2 
\wedge d\phi
=\frac{L^4\sinh^2\rho\sin\theta}{\sin^4\eta}
d\rho\wedge d\psi\wedge d\theta\wedge d\phi\,.
\end{equation}
The crucial point is that the conformal transformation yields a potential 
that differs from the one we used, by a component along the $y$ direction. 
This comes about because the $AdS$ isometries mix $y$ with the rest of 
the coordinates.

The two potentials are related by a gauge
transformation, $C_4'=C_4+d\Lambda _3$, with 
$\Lambda_3=-\frac{L^4 r_2}{2y^2}d\psi \wedge dr_2 \wedge d\phi$. 
It is important to note that were we to compute the WZ term with this 
$C_4'$, we would just get the same answer as in the DBI term 
(that is only the first term in (\ref{WZ-circle})). This would 
make the bulk action zero, as in the case of the straight line.

This is not surprising. Using $C_4'$ just amounts to doing the calculation 
of the straight line in a different coordinate system. The subtle difference 
between the line and the circle is related here to the choice of four-form 
potential. One could find a similar ambiguity in the string calculation of 
the Wilson loop if one would regularize the divergence differently and 
remove the wrong boundary term. Likewise, in the perturbative calculation, 
the conformal transformation from the line to the circle amounts to a 
singular gauge transformation that adds a finite piece to the propagator
\cite{Drukker:2000rr}. That is a very close analog to the gauge 
transformation generated by $\Lambda_3$ above.

\section{'t Hooft loop}

The D3-brane calculation and the matrix model agreed for all finite values 
of the parameter $\kappa$ at large $N$ (so both $N\gg1$ and 
$\kappa N\gg1$), which in turn implies that $\lambda \gg 1$.
So far the $AdS$ calculation was 
valid only in the range $1\ll\lambda\ll N$, where the string coupling 
is weak. We wish now to 
consider the case when $\lambda\gg N$, or strong coupling. To study 
it we will have to go to the S-dual theory, which is weakly coupled.

Under S-duality we replace $g_s$ with $1/\tilde g_s$, or 
$\lambda=16\pi^2N^2/\tilde\lambda$. We can express $\kappa$ in 
terms of the dual couplings as $\pi k/\sqrt{\tilde\lambda}$. We have 
to distinguish between two cases, when $1\ll\tilde\lambda\ll N$, 
or $N\ll\lambda\ll N^2$, we should perform the calculation in the 
S-dual $AdS$ space. When $\tilde\lambda\ll1$, or 
$\lambda\gg N^2$, this $AdS$ is highly curved, and instead we should 
use the dual gauge theory, which is now weakly coupled.

S-duality does not only change the coupling, but also the charges. 
Electric charge is replaced with magnetic, so the Wilson loop we are 
studying will be an 't Hooft loop \cite{'tHooft:1977hy} in the dual 
theory. 

\subsection{'t Hooft loop in $AdS$}

The standard prescription for calculating an 't Hooft loop in the 
$AdS$/CFT correspondence is, like for the Wilson loop, by use of a 
minimal surface with the substitution of the fundamental string 
by a D1-brane. Again we will look to replace this D1-brane by 
a D3-brane, which happens very naturally in the presence of background 
flux, and goes under the name of the Myers effect 
\cite{Myers:1999ps}. We do the calculation here for a circular 
loop, the straight line can be done in a similar fasion.

The construction will be the same as before, the only difference is the 
replacement of the electric field by its Hodge dual, a magnetic field on 
the $S^2$ of our favorite coordinate system (\ref{circle-metric}). 
Explicitly we shall take $F_{\theta\phi}=(k\sin\theta)/2$. 
Defined this way, $k$ is the number of D1-branes immersed in the 
D3-branes.

The WZ part of the action is identical to the 
electric case (\ref{WZ-circle}), while the DBI part is now
\begin{eqnarray}
S_{\rm DBI}
&=&T_{\rm D3}\int e^{-\Phi}\sqrt{\det(g+2\pi\alpha' F)}
\nonumber\\
&=&
2N\int d\rho\,d\theta\,
\frac{\sin\theta\sinh^2\rho}{\sin^4\eta}
\sqrt{\cos^2\eta(1+\eta'^2)\left(1+
\frac{\pi^2k^2\sin^4\eta}{\tilde\lambda\sinh^4\rho}
\right)}\,.
\end{eqnarray}
The solution to the equations of motion is, as before, 
$\sin\eta=\kappa^{-1}\sinh\rho$, now with 
$\kappa=\pi k/\sqrt{\tilde\lambda}$. On shell this part of the action 
is evaluated to be
\begin{equation}
S_{\rm DBI}
=4N(\kappa^2+\kappa^4)\int d\rho\,
\frac{1}{\sinh^2\rho}
=-4N(\kappa^2+\kappa^4)\coth\rho
\Bigg|_{\sinh\rho=\kappa\sin\eta_0}^{\sinh\rho=\kappa}\,.
\end{equation}
Again we should replace the coordinate $\eta$ with the conjugate 
momentum $p_\eta$ by adding the boundary term
\begin{equation}
S_{\rm boundary}=-\eta_0\int d\psi\, p_\eta
=-4N\frac{\kappa}{\eta_0}\,.
\end{equation}
This cancels one of the divergent terms in the DBI action. 
The other divergence, $4N\kappa^3/\eta_0$ will 
cancel against the divergent term in the WZ action.

Combining with the WZ term we find the final answer with the 
same functional form as before (\ref{final})
\begin{equation}
S_{\rm circle}=
-2N\left [\kappa\sqrt{1+\kappa^2}
+\sinh^{-1}\kappa\right]\,.
\end{equation}

There are some important differences from the electric case. While there 
we had to replace the field strength by its dual, for the case at hand the 
magnetic field is the correct variable counting the number of D1-branes 
dissolved in the D3-brane. Consequently the bulk action had the 
exact same linear divergence found when calculating the 't Hooft loop 
by means of D1-branes.

\subsection{'t Hooft loop in perturbation theory}

In the regime when $\lambda\gg N^2$, or $\tilde\lambda\ll 1$ 
we can no longer use the dual $AdS$, and instead have to study the 
weakly coupled dual gauge theory. The expectation value of an 't Hooft 
loop in four dimensions was never calculated, as there are some 
technical hurdles that are yet to be overcome. Still we will try to 
carry this calculation as far we can\footnote{Based in part on 
a collaboration between N.D. and N. Itzhaki \cite{sunny}.}.

We may try to extend the results of the matrix model for finite $N$ to this 
regime. In that case we need to evaluate the Laguerre polynomial in 
(\ref{laguerre}) at 
very large negative arguments, where it will be dominated by its highest 
exponent $L_n^k(x)\sim 1/n! (-x)^n$. This gives
\begin{equation}
\vev{V}=\frac{1}{N}L_{N-1}^1(-4N\kappa^2)\exp[2N\kappa^2]
\sim\frac{1}{N!}
\left(\frac{2\pi k}{\tilde g_{YM}}\right)^{2N-2}
\exp\left[\frac{2\pi^2k^2}{\tilde g_{YM}^2}\right]\,.
\label{matrix-tHooft}
\end{equation}
A very similar result is found from the Wilson loop and 't Hooft loop 
calculation in terms 
of the D3-brane. The result (\ref{final}) expanded for large $\kappa$ is
\begin{equation}
S=-2N\left(\kappa\sqrt{1+\kappa^2}+\sinh^{-1}\kappa\right)
\sim -2N\left(\kappa^2+\ln2\kappa+1/2+\cdots\right)\,.
\end{equation}
Switching again to the dual variables we find that the 't Hooft loop is 
given by
\begin{equation}
\vev{V}=\exp[-S]
\sim\left(\frac{e}{N}\right)^N
\left(\frac{2\pi k}{\tilde g_{YM}}\right)^{2N}
\exp\left[\frac{2\pi^2k^2}{\tilde g_{YM}^2}\right]\,.
\end{equation}

Unlike the Wilson loop calculation, where at large $N$ the $\sqrt\lambda$ 
behavior appeared, there seems to be no subtlety in taking the large $N$ 
limit, on the 't Hooft loop. The two results agree.

It is not too hard to explain the different factors in this expression 
in terms of a perturbative gauge theory calculation for $k=1$ 
\cite{sunny}.

The 't Hooft loop is a topological defect creating a magnetic source in the 
non-Abelian gauge theory. For the circular source it is useful to employ 
the coordinates $\rho$, $\psi$, $\theta$ and $\phi$ of (\ref{coords}) 
with $\eta=0$, so the flat space metric is
\begin{equation}
ds^2=\frac{R^2}{(\coth\rho-\cos\theta)^2}
\left[\frac{1}{\sinh^2\rho}(d\rho^2+d\psi^2)
+d\theta^2+\sin^2\theta d\phi^2\right]\,.
\end{equation}
In these coordinates the source is at $\rho=0$.

The expectation value of the 't Hooft loop should be given by the 
partition function in the background generated by this magnetic source 
(which in the supersymmetric case also carries scalar charge of the field 
we label $X^9$). At the 
classical level this magnetic field will sit in one $U(1)$ 
factor, and we can write the form of the electromagnetic field 
configuration.  It is given by
\begin{eqnarray}
X^9&=&
\frac{1}{2R}(\coth\rho-\cos\theta)\,,
\nonumber\\
A_\phi&=&\frac{1}{2}(\pm1-\cos\theta)\,,
\end{eqnarray}
where the sign choice in the gauge field corresponds to the two gauges 
covering the north or south pole of the sphere. The straight forward way 
to get this expression is by solving the magnetostatic problem for a 
monopole source along the circle (as well as the more familiar 
scalar). The Laplacean in our coordinate system is closely related to 
that on $AdS_2\times S^2$, and is easy to invert. 
Another way is to start with the straight line and do the conformal 
transformation to the circle. Finally one can just notice that 
this is the same solution as that of the DBI action with $X^9$ 
replacing $1/y$.

The classical action for this configuration includes two terms, from the 
gauge field and from the scalar. Both are divergent, but it is easy to 
regularize the integrals and extract a finite answer
\begin{equation}
S=\frac{1}{4\tilde g_{YM}^2}\int\sqrt{g}
\left(F^2+(\partial X^9)^2\right)
\sim -\frac{2\pi^2}{\tilde g_{YM}^2}\,.
\end{equation}
While at the technical level it is easy to subtract the divergence, it should 
really not exist in the first place. In all the other calculations the 
complete final result, including perhaps boundary terms is finite. This is 
true for 
the calculations of the Wilson loop as well as the 't Hoof loop in $AdS$ 
using either fundamental and D-strings or D3-branes. The same is true for 
the perturbative calculation of the Wilson loop. This is an indication that 
we are missing some terms in the action localized on the defect. At 
the classical level they will merely fix this divergence, hence we are 
able to continue without fully understanding them.

The leading behavior of the 't Hooft loop will be the exponent of minus 
this classical action, and indeed we see that this result agrees with the 
matrix model and D3-brane calculations. At the semi-classical level 
one has to quantize the theory around this background and evaluate the 
fluctuation determinant of all the fields. This calculation is naturally 
broken up into the zero mode contribution and that of the massive 
fluctuations. The calculation of the latter is beyond the 
scope of the present paper, and we will only comment on the zero-mode 
determinant.

The classical solution was an Abelian ansatz living in a $U(1)$ factor 
within $U(N)$, breaking the gauge symmetry to $U(N-1)\times U(1)$. This 
breaking results in $2N-2$ zero modes  each contributing a factor of 
$1/\tilde g_{YM}$ to the one-loop determinant. More precicely, those 
zero modes parameterize a coset manifold and their 
contribution is just the volume of this coset.

The volume of this manifold can be calculated in similar ways to that done 
for the zero modes of instantons (see for example 
\cite{Bernard:1979qt}). If one normalizes the generators of $U(N)$ 
such that $\Tr T^aT^b=\delta^{ab}/2$, the volume of $U(N)$ can 
be written in terms of the volume of $U(N-1)$ and that of the $2N-1$ 
dimensional unit sphere ($2\pi^N/(N-1)!$) as
\begin{equation}
V(U(N))=2^{2N-3/2}V(S^{2N-1})V(U(N-1))\,.
\end{equation}
The subgroup that is preserved by our ansatz is the product of $U(N-1)$ and 
a $U(1)$ of radius $2\pi\sqrt2$. 
The result one gets is
\begin{equation}
\frac{(4\pi)^{N-1}}{(N-1)!\,\tilde g_{YM}^{2N-2}}\,.
\end{equation}

Together the classical action and the zero modes give the answer
\begin{equation}
\vev{V}\sim \frac{1}{(N-1)!}
\left(\frac{2\sqrt\pi}{\tilde g_{YM}}\right)^{2N-2}
\exp\left[\frac{2\pi^2}{\tilde g_{YM}^2}\right]\,.
\end{equation}
This result is remarkably close to (\ref{matrix-tHooft}). The discrepancy 
in the powers of $\pi$ may be fixed by the determinant of the massive 
fluctuations. The source of the missing power of $N$ is unclear and could 
be related to the normalization of the 't Hooft operator, if it is dual 
to the Wilson loop defined without the factor of $1/N$ before the trace.

We should note that the calculations in the entire paper 
are concerned with $U(N)$ gauge theory. Everything can be generalized 
to $SU(N)$ in a reasonably simple manner. This is of importance 
for the 't Hooft loop, since it is a well defined operator only in the latter 
case. The resulting modifications are the rescaling of the action 
by a factor of $(N-1)/N$ and the decrease in the zero mode determinant 
by a factor of $N-1$.

While this perturbative calculation captured some of the salient 
features of the 't Hooft loop expectation value as evaluated from the 
matrix model and $AdS$, it suffers from some 
serious flaws. To do any better one may have to add to the gauge 
theory new degrees of freedom living along the loop, giving 
a defect CFT, like in \cite{DeWolfe:2001pq}.

\section{Discussion}

We have found D3-brane solutions in $AdS_5$ that carry electric flux and 
end along a curve on the boundary. This is the correct prescription for 
calculating the expectation value of a Wilson loop multiply wrapped 
around that curve when the multiplicity of the loop $k$ is big.

The expectation value of the Wilson loop calculated this way includes 
sereval different parts, the bulk contribution comprising of the DBI and 
the WZ pieces, and 
in addition the boundary terms were crucial to finding the right answer. 
One has to replace the coordinate transverse to the $AdS$ boundary with 
its conjugate momentum, and apply the same procedure to the gauge 
field (in the electric case).

Each of the different terms in the action had a linear divergence, but 
they all canceled in the final answer. In the case of the straight line 
the final result was zero, while for the circle there was a finite remnant.

In the circular example, where the answer is nontrivial, we found the 
D3-brane solution to reproduce correctly the string result 
$k\sqrt\lambda$ as well as an infinite series of corrections. The first 
of those $k^3\lambda^{3/2}/96N^2$ is seen as the first string-loop 
correction to the above result. The full D3-brane solution includes 
corrections to all orders in $1/N$ that are leading at large $\lambda$. 

Thus we find that
\begin{quotation}
{\em
The action of the D3-brane carrying electric flux and ending along the 
loop on the boundary captures correctly the action of the analogous 
fundamental strings as well as the entropy of summing over 
semi-classical string surfaces of \underline{all} genus.
}
\end{quotation}
Furthermore in this case we were able to compare the results to 
the Gaussian matrix model and found perfect agreement.

It is amusing to compare the $k$ dependence we found for this Wilson loop
in a conformal theory, and the tension of $k$-strings in supersymmetric
confining theories, as given by the Douglas-Shenker formula
\begin{equation}
\sigma_k=N\Lambda^2\sin \frac{\pi k}{N}=\Lambda ^2 \pi k-
\frac{\Lambda ^2\pi^3}{3!}\frac{k^3}{N^2}+\dots
\end{equation}

In both cases the leading term scales with $k$, and more interestingly,
the first correction in a large $N$ expansion goes like $k^3/N^2$. This
$k^3$ scaling is not expected a priori, and it would be nice to develop
a intuitive understanding of it (see \cite{armoni} for a heuristic
picture of this scaling for $k$-strings).

It would be very interesting to study other Wilson loops using this 
prescription. Probably the most interesting example would be the pair 
of anti-parallel lines. The standard string result for a pair of $k$ 
coincident lines at a distance $r$ gives the effective potential 
$V(r)=-4k\pi^2\sqrt{2\lambda}/(\Gamma(1/4)^4r)$. 
Finding the D3-brane would allow us to deduce 
the corrections to the effective multi-string tension beyond this 
leading behavior. It is natural to expect that the correction will again 
be of order $k^3\lambda^{3/2}/N^2$, but one would have to do the
full calculation to find the numerical coefficient.

We expect the general features of the D3-brane solution to carry over to all 
other Wilson loops including the parallel lines. In the UV any smooth 
loop looks like a straight line, thus the D3-brane will have the same 
asymptotic form, with the geometry approaching $AdS_2\times S^2$. 
All the divergences come from the UV and will cancel as in the two 
examples studied here, but a finite contribution, dependent on the shape 
of the curve, will remain.

Let us note that while we got those non-planar contributions to the Wilson 
loop quite easily for the circle 
by using the D3-branes, there should be other ways of calculating them. 
One direction is to follow \cite{Berenstein:1999ij} and calculate the 
exchange of supergravity fields between different parts of the 
string surface. Another approach would be to follow the argument 
given in \cite{Drukker:2000rr} and reviewed above for the leading 
power of $\lambda$ in the first $1/N^2$ correction. The factor of 
$1/96$ should come out of the ratio of volumes of the moduli spaces of 
degenerate genus one Riemann surfaces divided by the genus zero case.

Another interesting direction is to look at other terms in the D3-brane 
action. There are curvature corrections, which would correspond to 
terms that are subleading in $\lambda$ to the ones we found 
(or subleading in $N$, if we keep $\kappa$ constant). Those terms 
can also be calculated from the string solution by using world-sheet 
techniques, but while the prescription for performing calculation is 
straight-forward, the calculation itself is pretty hard 
\cite{Drukker:2000ep}. 
It may, therefore, be advantageous to use the D3-brane for this 
purpose, and again compare the result to the matrix model. 

Other terms in the action correspond to D-instanton contributions. 
Those should agree with the instanton correction to the perturbative 
result \cite{Bianchi:2002gz} (which are not captured by the matrix 
model).

From the supergravity perspective the difference between a multiply 
wrapped Wilson loop and an operator with more than one trace is 
subtle. They are all described by $k$ coincident fundamental strings, 
where the difference is whether the strings are independent, or 
connected along branch points to a single Riemann surface 
\cite{Gross:1998gk}. From the comparison to the matrix model it 
seems like the single D3-brane solution corresponds to the single 
trance operator in the gauge theory. It's natural to guess that in the 
general case we should have one D3-brane for each trace in the Wilson 
loop operator (or each disconnected string surface). That system may then 
be an excellent laboratory for studying the non-Abelian generalization of 
the DBI action.

It is quite remarkable how this Gaussian matrix model 
\cite{Drukker:2000rr} captures interesting string phenomena. It is 
successful in reproducing the $AdS$ calculation of the Wilson loop 
using classical strings and all the higher genus corrections to it, or 
the D3-brane solution. The matrix model still includes many more 
terms on top of the ones studied here, and a few more of them 
will be compared to $AdS$ in \cite{we}.

\section*{Acknowledgments}
We would like to thank Ofer Aharony, Jan Ambj{\o}rn, Tom Banks, 
Micha Berkooz, Sunny Itzhaki, Yuri Makeenko, David Mateos, 
John McGreevy, Rob Myers, Joan Sim\'on and Kostas Skenderis 
for interesting conversations. We are also grateful to Soo-Jong Rey 
for pointing out the special relevance of \cite{rey-wl}. 
N.D. would like to thank the Weizmann Institute, where this collaboration 
was initiated, for its hospitality as well as the 
Aspen Center for Physics and the University of Amsterdam. 
B.F. is grateful for the hospitality of the Aspen Center for Physics 
and the University of California at Santa Cruz.


\begin{thebibliography}{20}
%%%%%%%%%%%%%%%%%%%%%%%%%%%
\addtolength{\parskip}{-1ex}

\bibitem{ads/cft}
J.~M.~Maldacena,
``The large $N$ limit of superconformal field theories and supergravity,''
Adv.\ Theor.\ Math.\ Phys.\  {\bf 2}, 231 (1998)
[Int.\ J.\ Theor.\ Phys.\  {\bf 38}, 1113 (1999)]
[arXiv:hep-th/9711200].
%%CITATION = HEP-TH 9711200;%%

\bibitem{rey-wl}
S.~J.~Rey and J.~T.~Yee,
``Macroscopic strings as heavy quarks in large $N$ gauge theory and anti-de
Sitter supergravity,''
Eur.\ Phys.\ J.\ C {\bf 22}, 379 (2001)
[arXiv:hep-th/9803001].
%%CITATION = HEP-TH 9803001;%%

\bibitem{maldacena-wl}
J.~M.~Maldacena,
``Wilson loops in large $N$ field theories,''
Phys.\ Rev.\ Lett.\  {\bf 80}, 4859 (1998)
[arXiv:hep-th/9803002].
%%CITATION = HEP-TH 9803002;%%

\bibitem{McGreevy:2000cw}
J.~McGreevy, L.~Susskind and N.~Toumbas,
``Invasion of the giant gravitons from anti-de Sitter space,''
JHEP {\bf 0006}, 008 (2000)
[arXiv:hep-th/0003075].
%%CITATION = HEP-TH 0003075;%%

\bibitem{Grisaru:2000zn}
M.~T.~Grisaru, R.~C.~Myers and O.~Tafjord,
``SUSY and Goliath,''
JHEP {\bf 0008}, 040 (2000)
[arXiv:hep-th/0008015].
%%CITATION = HEP-TH 0008015;%%

\bibitem{Hashimoto:2000zp}
A.~Hashimoto, S.~Hirano and N.~Itzhaki,
``Large branes in $AdS$ and their field theory dual,''
JHEP {\bf 0008}, 051 (2000)
[arXiv:hep-th/0008016].
%%CITATION = HEP-TH 0008016;%%

\bibitem{Callan}
C.~G.~.~Callan and J.~M.~Maldacena,
``Brane dynamics from the Born-Infeld action,''
Nucl.\ Phys.\ B {\bf 513}, 198 (1998)
[arXiv:hep-th/9708147].
%%CITATION = HEP-TH 9708147;%%

\bibitem{Gibbons}
G.~W.~Gibbons,
``Born-Infeld particles and Dirichlet p-branes,''
Nucl.\ Phys.\ B {\bf 514}, 603 (1998)
[arXiv:hep-th/9709027].
%%CITATION = HEP-TH 9709027;%%

\bibitem{Gross:1998gk}
D.~J.~Gross and H.~Ooguri,
``Aspects of large $N$ gauge theory dynamics as seen by string theory,''
Phys.\ Rev.\ D {\bf 58}, 106002 (1998)
[arXiv:hep-th/9805129].
%%CITATION = HEP-TH 9805129;%%

\bibitem{douglas}
M.~R.~Douglas and S.~H.~Shenker,
``Dynamics of $SU(N)$ supersymmetric gauge theory,''
Nucl.\ Phys.\ B {\bf 447}, 271 (1995)
[arXiv:hep-th/9503163].
%%CITATION = HEP-TH 9503163;%%

\bibitem{hanany}
A.~Hanany, M.~J.~Strassler and A.~Zaffaroni,
``Confinement and strings in M{QCD},''
Nucl.\ Phys.\ B {\bf 513}, 87 (1998)
[arXiv:hep-th/9707244].
%%CITATION = HEP-TH 9707244;%%

\bibitem{herzog}
C.~P.~Herzog and I.~R.~Klebanov,
``On string tensions in supersymmetric $SU(M)$ gauge theory,''
Phys.\ Lett.\ B {\bf 526}, 388 (2002)
[arXiv:hep-th/0111078].
%%CITATION = HEP-TH 0111078;%%

\bibitem{hirosi}
N.~Drukker, D.~J.~Gross and H.~Ooguri,
``Wilson Loops and Minimal Surfaces,''
Phys.\ Rev.\  {\bf D60}, 125006 (1999)
[hep-th/9904191].
%%CITATION = HEP-TH 9904191;%%

\bibitem{Berenstein:1999ij}
D.~Berenstein, R.~Corrado, W.~Fischler and J.~Maldacena,
``The Operator Product Expansion for Wilson Loops and Surfaces in the
Large $N$ Limit,''
Phys.\ Rev.\  {\bf D59}, 105023 (1999)
[hep-th/9809188].
%%CITATION = HEP-TH 9809188;%%

\bibitem{Bianchi:2002gz}
M.~Bianchi, M.~B.~Green and S.~Kovacs,
``Instanton corrections to circular Wilson loops in $\cN = 4$ supersymmetric
Yang-Mills,''
JHEP {\bf 0204}, 040 (2002)
[arXiv:hep-th/0202003].
%%CITATION = HEP-TH 0202003;%%

\bibitem{Mikhailov:2002ya}
A.~Mikhailov,
``Special contact Wilson loops,''
arXiv:hep-th/0211229.
%%CITATION = HEP-TH 0211229;%%

\bibitem{erickson}
J.~K.~Erickson, G.~W.~Semenoff and K.~Zarembo,
``Wilson loops in $\cN = 4$ supersymmetric Yang-Mills theory,''
Nucl.\ Phys.\ B {\bf 582}, 155 (2000)
[arXiv:hep-th/0003055].
%%CITATION = HEP-TH 0003055;%%

\bibitem{Drukker:2000rr}
N.~Drukker and D.~J.~Gross,
``An exact prediction of $\cN = 4$ SUSYM theory for string theory,''
J.\ Math.\ Phys.\  {\bf 42} (2001) 2896
[arXiv:hep-th/0010274].
%%CITATION = HEP-TH 0010274;%%

\bibitem{Gauntlett}
J.~P.~Gauntlett, C.~Koehl, D.~Mateos, P.~K.~Townsend and M.~Zamaklar,
``Finite energy Dirac-Born-Infeld monopoles and string junctions,''
Phys.\ Rev.\ D {\bf 60}, 045004 (1999)
[arXiv:hep-th/9903156].
%%CITATION = HEP-TH 9903156;%%

\bibitem{bachas}
C.~Bachas and M.~Petropoulos,
``Anti-de-Sitter D-branes,''
JHEP {\bf 0102}, 025 (2001)
[arXiv:hep-th/0012234].
%%CITATION = HEP-TH 0012234;%%

\bibitem{Skenderis:2002vf}
K.~Skenderis and M.~Taylor,
``Branes in $AdS$ and pp-wave spacetimes,''
JHEP {\bf 0206}, 025 (2002)
[arXiv:hep-th/0204054].
%%CITATION = HEP-TH 0204054;%%

\bibitem{Witten}
E.~Witten,
``Anti-de Sitter space and holography,''
Adv.\ Theor.\ Math.\ Phys.\  {\bf 2}, 253 (1998)
[arXiv:hep-th/9802150].
%%CITATION = HEP-TH 9802150;%%

\bibitem{Emparan:2001wn}
R.~Emparan and H.~S.~Reall,
``A rotating black ring in five dimensions,''
Phys.\ Rev.\ Lett.\  {\bf 88}, 101101 (2002)
[arXiv:hep-th/0110260].
%%CITATION = HEP-TH 0110260;%%

\bibitem{'tHooft:1977hy}
G.~'t Hooft,
``On The Phase Transition Towards Permanent Quark Confinement,''
Nucl.\ Phys.\ B {\bf 138}, 1 (1978).
%%CITATION = NUPHA,B138,1;%%

\bibitem{Myers:1999ps}
R.~C.~Myers,
``Dielectric branes,''
JHEP {\bf 9912}, 022 (1999)
[arXiv:hep-th/9910053].
%%CITATION = HEP-TH 9910053;%%

\bibitem{sunny}
N.~Drukker and N.~Itzhaki, unpublished.

\bibitem{Bernard:1979qt}
C.~Bernard,
``Gauge zero modes, instanton determinants, and quantum-chro\-modynamic
calculations,''
Phys.\ Rev.\ D {\bf 19}, 3013 (1979).
%%CITATION = PHRVA,D19,3013;%%

\bibitem{DeWolfe:2001pq}
O.~DeWolfe, D.~Z.~Freedman and H.~Ooguri,
``Holography and defect conformal field theories,''
Phys.\ Rev.\ D {\bf 66}, 025009 (2002)
[arXiv:hep-th/0111135].
%%CITATION = HEP-TH 0111135;%%

\bibitem{armoni}
A.~Armoni and M.~Shifman,
``Remarks on stable and quasi-stable $k$-strings at large $N$,''
Nucl.\ Phys.\ B {\bf 671}, 67 (2003)
[arXiv:hep-th/0307020].
%%CITATION = HEP-TH 0307020;%%

\bibitem{Drukker:2000ep}
N.~Drukker, D.~J.~Gross and A.~A.~Tseytlin,
``Green-Schwarz string in $AdS_5\times S^5$: Semiclassical 
partition  function,''
JHEP {\bf 0004}, 021 (2000)
[arXiv:hep-th/0001204].
%%CITATION = HEP-TH 0001204;%%

\bibitem{we}
N.~Drukker and B.~Fiol, In preparation.


\end{thebibliography}
\end{document}